# Role of Surface Gravity Waves in Aquaplanet Ocean Climates

J. H. P. Studholme[a#], M. Yu. Markina[b#*] and S. K. Gulev[bc]

[#]These authors are sharing lead author status

[a]Yale University, New Haven, CT. USA

[b]Shirshov Institute of Oceanology, Russian Academy of Science, Moscow, Russia

[c]Lomonosov Moscow State University, Moscow, Russia

[*]Present affiliation: University of Oxford, UK.

Corresponding author: Josh Studholme (josh.studholme@gmail.com), https://orcid.org/0000-0003-0704-0286

Corresponding author: Rita Markina (markina@sail.msk.ru), https://orcid.org/0000-0003-3563-8242

**Key points:**

1. Idealized oceanic climates forced by ranging atmospheric regimes following equator-to-pole thermal gradient perturbations are investigated.

2. Wind-forced surface gravity waves deepen the mixed layer, increase mixed layer vertical momentum diffusivity and dampen surface currents.

3. The consistency of the effects of waves on ocean dynamics and stratification across cooler/warmer aquaplanet climates is examined.

**Manuscript length: 25/25 PU – 7.5k words (15 PU), 10 plots (10 PU)**

**Plain Language Summary**

In this study, we investigate how wind-driven ocean surface waves impact the ocean properties using a combination of numerical models. We simulate atmospheric and oceanic circulation on an aquaplanet (analogous to the Earth without continents). This allows us to study the ocean response to different regimes of large-scale atmospheric circulation which are set simply by changes in the gradient of sea surface temperature between the tropics and the poles. We show that ocean surface waves contribute to vertical mixing and the surface stress that the atmosphere exerts upon the ocean. This is manifested as deepening the mixed layer and dampening of the upper ocean currents which reduces the strength of oceanic meridional overturning circulations in subtropics and midlatitudes. Surface waves contribute to upper ocean mixing both directly and via induced changes in buoyancy and wind-induced currents. Including surface gravity wave parameterisation in the ocean model increases climatological mixed layer depth by ~10%, an effect which is found to be robust under different climate states.

**Abstract**

We present a set of idealised numerical experiments of a solstitial aquaplanet ocean and examine the thermodynamic and dynamic implications of surface gravity waves (SGWs) upon its mean state. The aquaplanet's oceanic circulation is dominated by an equatorial zonal jet and four Ekman driven meridional overturning circulation (MOC) cells aligned with the westerly atmospheric jet streams and easterly trade winds in both hemispheres. Including SGW parameterization (representing modulations of air-sea momentum fluxes, Langmuir circulation and Stokes-Coriolis force) increases mixed layer vertical momentum diffusivity by ~40% and dampens surface momentum fluxes by ~4%. The correspondingly dampened MOC impacts the oceanic density structure to 1 km depth by lessening the large-scale advective transports of heat and salt, freshening the equatorial latitudes (where evaporation minus precipitation [E-P] is negative) and increasing salinity in the subtropics (where E-P is positive) by ~1%. The midlatitude pycnocline in both hemispheres is deepened by the inclusion of SGWs. Including SGWs into the aquaplanet ocean model acts to increase mixed layer depth by ~10% (up to 20% in the wintertime in midlatitudes), decrease vertical shear in the upper 200 m and alter local midlatitude buoyancy frequency. Generally, the impacts of SGWs upon the aquaplanet ocean are found to be consistent across cooler and warmer climates. We suggest that the implications of these simulations could be relevant to understanding future projections of SGW climate, exoplanetary oceans, and the dynamics of the Southern Ocean mixed layer.

**Introduction**

Wind-driven ocean surface gravity waves (SGWs) play a crucial role in air-sea interaction by actively modulating surface heat, momentum and mass exchanges (Cavaleri, 2012). Two of the most significant mechanical impacts of these waves are (1) inducing and amplifying mixing in the ocean surface boundary layer (OSBL, Craig and Banner, 1994; Grant and Belcher, 2009; Belcher et al., 2012; Wu et al., 2015; Li et al., 2016, Buckingham et al. 2019) and (2) mediating air-sea momentum fluxes and thus the wind-driven oceanic circulations (Janssen, 1991; Gulev and Hasse, 1998; Kukulka et al., 2007; Troitskaya et al., 2012). Implementing parameterized SGW impacts into Ocean General Circulation Models (OGCMs) leads to better representation of mixed layer depth (MLD; Babanin et al., 2009; Fan and Griffies, 2014; Breivik et al., 2015; Li et al., 2016, Walsh et al., 2017), ocean currents (Sullivan and McWilliams, 2010; McWilliams and Fox-Kemper, 2013; Villas Bôas et al., 2020) and ocean heat content (Stoney et al., 2018, Shimura et al., 2020). Improved representation of air-sea momentum fluxes results in global-scale climate bias reductions in both the ocean and atmosphere (Garfinkel et al., 2011). Thus, despite being a fine-scale phenomenon that is subgrid scale in contemporary OGCMs, SGWs play an important role at the large scale (McWilliams, 2016). Much of this role remains unexplored in relation to large-scale climate variability and change (Fan and Griffies, 2014).

SGWs impact the upper ocean because they project their influence down throughout both the Ekman layer and the mixed layer. Stokes drift, net drift following wave propagation, decays with a depth scale of $\lambda/4\pi$ ($\lambda$ – wavelength). So, a monochromatic wave with period 12 s projects its influence down ~18 m. However, instability arising from the interaction between Stokes drift and local vorticity gives rise to Langmuir turbulence and associated downwelling jets that penetrate throughout the Ekman layer to ~100 m depth (Polton and Belcher, 2007; Belcher et al., 2012; D'Asaro et al., 2014). Under certain conditions Langmuir turbulence creates a regime different from regular upper ocean shear driven turbulence. This arises when the non-dimensional turbulent Langmuir number $La_t$ < 0.5; $La_t = \sqrt{u_*/u_S^0}$, $u_*$ is friction velocity attributed to total wind stress, $u_S^0$ is surface Stokes drift (McWilliams et al., 1997). This process represents a separate source of turbulent kinetic energy (TKE) in the OSBL (Grant and Belcher, 2009).

The effects from SGW-induced mixing, in particular Langmuir turbulence, are parameterized into OGCMs via turbulent diffusion in KPP vertical mixing schemes (Reichl et al., 2016; Li et al., 2017, 2019), energetic-based planetary boundary layer (ePBL) parameterization of shear and convective turbulence (Reichl and Li, 2019) and as an additional source term of TKE in k-ε models (Axell et al., 2002). Efforts have also been made to develop second-moment closure models of Langmuir turbulence (Harcourt, 2013; 2015) however these are yet to be implemented in OGCMs. SGW models have been coupled with the following climate models: FIO-ESM (Qiao et al., 2013), ACCESS (Bi et al., 2013), GFDL CM2M (Fan and Griffies, 2014) and CESM2 (Li et al., 2016).

Many of the most prominent atmospheric phenomena driving SGW variability at the large-scale, in particular the atmospheric jet streams and trade winds, are features of the zonal-mean winds (Fan et al. 2013, 2014; Markina et al., 2019; Sharmar et al., 2021). The variability and dynamics of these phenomena have long been studied using the aquaplanet idealization (e.g., Hayashi and Sumi, 1986; Medeiros et al., 2016). In this approach the many complex basin-specific details are suspended in favour of unveiling generalised behaviour and intra-model differences (Blackburn and Hoskins, 2013). In this way, aquaplanets form a core part of the modelling hierarchy as a link between simple conceptual models and planetary scale, full physics simulations (Maher et al., 2019). They provide an essential role in benchmarking modelling approaches (Neale and Hoskins, 2000) and advancing conceptual insight into the large-scale circulation (Shaw, 2019).

To our knowledge, SGWs have not yet been studied on an aquaplanet, likely because the approximation of zonal symmetry is less applicable to the oceanic circulation than it is to atmospheric dynamics. However, dynamical aquaplanet ocean models have been applied to study significant climate problems, particularly poleward heat transport (Smith et al., 2006; Marshall et al., 2007; Enderton and Marshall, 2009, Salameh et al., 2018). Most widely they are used to study exoplanet habitability and ocean dynamics (Ferreira et al., 2014; Hu and Yang, 2014; Cullum et al., 2016; Fujii et al., 2017; Kane et al., 2018; Yang et al., 2019; Del Genio et al., 2019; Checlair et al., 2019; Salazar et al., 2020; Yue and Yang, 2020; Yang et al., 2020). We propose that SGWs are relevant to exoplanet ocean dynamics as they are on Earth.

On the real Earth, many of the SGW trends found in the contemporary period (1980-2019) and those projected for the twenty-first century are directly related to trends in the zonal-mean winds (Fan et al., 2013, 2014; Young and Ribal, 2019; Sharmar et al., 2021). The strongest SGW trends are between 40 and 60°S, i.e. the Southern Ocean region (Young and Ribal, 2019; Sharmar et al., 2021), a region where traditional OGCMs, that neglect SGWs, tend to underestimate oceanic MLD (Sallee et al., 2013; Huang et al., 2014) and where MLD is poorly explained by contributions from shear and buoyancy alone (Panassa et al., 2018). This suggests that SGWs might be significant to the Southern Ocean mixed layer, which is a region where zonal symmetry is a reasonable first-order simplification (Smith et al., 2006).

Thus, aquaplanets provide a simplified context for understanding SGW's active role in a dynamic ocean and may be useful in understanding projections of twenty-first century SGW climate, exoplanet climates and the Southern Ocean mixed layer. Given this, we adopt the following research questions:

(1) *How does including surface wave processes impact an OGCM aquaplanet ocean?*

(2) *How do these simulated wave impacts change across different climates*?

Section 2 details our modelling approach. Section 3 examines a control climate with and without SGW parameterization and Section 4 explores cooler and warmer climates than the control simulation, again with and without SGWs. The paper ends with a synthesis of the results and a very brief further discussion of these simulations' relevance.

## 2. Methods

### 2.1 Model Formulation and Experiments

We conduct numerical experiments of an Earth-like aquaplanet using an Atmospheric General Circulation Model (AGCM), an Oceanic GCM coupled with a sea-ice model (OGCM) and an ocean surface Spectral Wave Model (SWM). The planetary size, mass, rotation rate and other celestial parameters all match a typical Earth model configuration. The AGCM is forced from below by surface enthalpy fluxes computed to correspond to a pre-defined zonally-symmetric sea surface temperature (SST) profile. From above, the solar radiative flux corresponds to terrestrial solstitial conditions achieved by fixing the orbital configuration on day 200 of Earth's annual cycle. We run this AGCM setup and use the simulated 10 m winds as forcing for the SWM. The resulting atmospheric and SGW fields are used to force the OGCM.

Specifically, the OGCM is forced by winds, temperature and specific humidity at 10 m height, longwave and shortwave radiative fluxes, sea level pressure and precipitation from the AGCM, in experiments with SGWs it also uses Stokes drift and the ratio between atmospheric stress ($\tau_{\text{air-side}}$) and the net stress going into the waves at the ocean surface ($\tau_{\text{water-side}}$) from the SWM (see Section 2.2). We adopt this approach, in contrast to an interactively coupled configuration, since we are interested in understanding the gross role played by SGWs in the formation of the mean upper-ocean state. Although, there is clearly equal merit in a complementary study using a fully coupled setup to compute the net role.

### 2.1.1. Atmospheric model

The atmospheric model is constructed out of the Isca framework (Vallis et al., 2018). It solves the primitive equations with an Eulerian spectral dynamical core configured for T42 (~2.8°) horizontal resolution over 41 unevenly-spaced vertical levels spanning from approximately 10 m to ~74 km. The spacing is approximately 50 m near the surface (to a height of 900 hPa), 500 m throughout the troposphere and 3 km in the stratosphere and beyond (spacing increases with height). We use essentially the same Isca setup as Thomson and Vallis (2018a, b). Radiative transfer is done using the Rapid Radiative Transfer code (RRTM; Clough et al., 2005), the boundary layer formulation is based on Monin-Obukhov similarity, and atmospheric convection is parameterized following the Betts-Miller scheme (Betts and Miller, 1986). RRTM uses a prescribed time-invariant ozone distribution that is zonally and hemispherically symmetric (Jucker and Gerber, 2017). The AGCM's surface energy budget is closed using a slab ocean of constant depth in which the time-varying horizontal heat transport is dynamically controlled through a '$Q$-flux' governed by the temperature equation:

$$c \frac{\partial T}{\partial t} = SW + LW - SEF + \nabla \cdot Q, \qquad (1)$$

where $c$ is the slab's heat capacity, $T$ is temperature, and $t$ is time. $SW$, $LW$ and $SEF$ are the net shortwave radiative, net longwave radiative, and the sum of sensible and latent heat fluxes. The $Q$-flux is calculated according to Russell et al. (1985) in such a way as to prevent the free running Isca mixed-layer SSTs drifting from the prescribed SST forcing field. Ocean albedo is set to 0.25, roughness length is $2\times10^{-4}$ m, the Isca MLD is 20 m and evaporative resistance is 0.7 following Thomson and Vallis (2018 a,b). The model timestep is 300 seconds for dynamics and 1800 seconds for the radiation code.

2.1.2. Ocean model

The OGCM simulations are done using NEMO v3.6 (Madec et al., 2016) coupled with the sea ice model LIM v3.6 (Rousset et al., 2015). NEMO is run on a regular grid with fixed horizontal resolution of 2° and 31 vertical levels unevenly distributed in depth. In the first 120 meters of the ocean the vertical resolution is ~10 m, from there to 1000 m depth it steadily increases from 15 to 150 m. Below 1000 m to the ocean bottom the vertical resolution is ~500 m. The planetary ocean depth is taken as 5000 m and implicit bottom friction is applied at the bottom of the ocean. To avoid the problem of mass convergence at polar singularities, we include small polar islands. The ocean circulation is therefore computed over a quasi-global domain from 80 °S to 80 °N. We adopt free-slip boundary conditions around these polar islands.

Air-sea momentum fluxes in NEMO are defined as:

$$\tau_{total} = \rho_a C_D |\Delta \vec{U}| \Delta \vec{U}, \qquad (2)$$

where $\Delta \vec{U} = \vec{U}(z_u) - \vec{U}_0$ is the vector difference between the wind at height $z_u$ (in our case, 10 m) and the surface current ($\vec{U}_0$), $\rho_a$ is air density, and $C_D$ is a drag coefficient which itself is a function of atmospheric stability and wind speed. The corresponding air-sea enthalpy fluxes into the ocean are computed by NEMO from the AGCM fields using the standard NCAR bulk formulae (Large and Yeager, 2004). In this way, we do not let the free-running simulated OGCM SSTs influence the surface fluxes into the ocean.

The ocean model is initialised from a state of rest and an analytical temperature profile similar to the climatological annual-mean zonal-mean tropical Pacific Ocean with constant salinity everywhere of 35.5 PSU. It is then spun up into a statistically steady state by running it for 1000 years forced by the constant atmospheric forcing of the 3-year average of the AGCM runs. No sea ice is initialised in the OGCM but during spin-up it develops in the SH but not the NH (see Supplementary Figure S1). Langmuir parameterisation is turned off when sea ice concentration is 0.25 which is at 48°S in the TM300 experiment (see Supplementary Figure S2). Momentum flux, including when modified by wave effects, is scaled by sea ice concentration. Sea ice concentration reaches unity at 68°S. We do not consider the ice-SGW interactions since the sea ice model is not coupled to the SGW model.

In computing lateral momentum diffusion, we set the value of lateral viscous velocity to 0.2 ms$^{-1}$ (c.f. with NEMO's ORCA2 value of 0.1 ms$^{-1}$ for a configuration with Earth's bathymetry). The ORCA2 value is tuned specifically to optimise representation of Earth's western boundary currents. In the aquaplanet ocean, the absence of meridional boundaries allows for strong zonal flows, the increase in lateral viscous velocity is done to satisfy the Courant-Friedrichs-Lewy stability criterion and reduce grid scale noise in these zonal jets. We use an eddy-induced velocity parameterization which introduces additional tracer advection in combination with the isoneutral diffusion of tracers with space-time variable coefficients. Throughout this analysis, mixed layer depth is estimated based on the turbocline depth criterion. This is the depth that the vertical eddy diffusivity coefficient determined by vertical physics alone (not the isopycnal component) falls below a given value of 0.05 m$^2$ s$^{-1}$ (Brodeau and Koenigk 2016).

### 2.1.3. Spectral wave model

The SGWs are simulated with WAVEWATCH III v5.16 (hereafter referred to as WW3; WW3DG, 2016) based on the wave action balance equation:

$$\frac{\partial A}{\partial t} + \frac{\partial \omega}{\partial k}\frac{\partial A}{\partial x} + \frac{\partial \omega}{\partial x}\frac{\partial A}{\partial k} = S_{in} + S_{ds} + S_{nl}, \qquad (3)$$

where $A$ is wave action density - itself a function of the Cartesian horizontal coordinate $x$, frequency $\omega$ and wavenumber $k$. Right-hand side represents energy input due to winds ($S_{in}$), energy dissipation ($S_{ds}$) and nonlinear wave-wave interactions ($S_{nl}$). The wave field is computed over the quasi-global oceanic domain from 80 °S to 80 °N with a horizontal resolution of 2.8° consistent with the resolution of the atmospheric integrations. We use the ST4 parameterization for wave energy input and dissipation (Ardhuin et al., 2010) and the Discrete Interaction Approximation (DIA) scheme for nonlinear wave interactions (Hasselmann and Hasselmann, 1985). As in the OGCM, ocean depth is constant and set everywhere to 5000 m. The model integration time step is 20 minutes.

### 2.1.4. Model experiments

Simulations are done for 1080 days (bar the 1,000 year OGCM spin-up) with the first 360 days being taken as the full system's spin-up period which is discarded from all subsequent calculations. This leaves 720 days or 24 months for consideration. Again, the atmosphere's lower boundary corresponds to a zonally-symmetric time-invariant SST field. We study five forcing profiles corresponding to perturbed tropical SSTs (see Figure 7a). These profiles are of the form originally introduced by Neale and Hoskins (2000; referred to there as $Q_{obs}$) and used in other aquaplanet experiments (e.g. Fedorov et al., 2019). They follow these functional forms:

$$T_s(\varphi) = \begin{cases} T_{min} + (T_{max} - T_{min})\left[1 - \frac{1}{2}(sin^2(\alpha) + sin^4(\alpha))\right]; & -60 + \varphi_{max} < \varphi < 60 + \varphi_{max} \\ T_{min}; & otherwise \end{cases} \quad (4)$$

with $\alpha = \frac{3(\varphi - \varphi_{max})}{2}$, $T_s$, $T_{min}$ and $T_{max}$ as SST, global minimum and maximum SST respectively (units K), $\varphi$ as latitude (units °N) and $\varphi_{max}$ is the latitude of $T_{max}$.

We define and study a control oceanic simulation without SGW parameterization and a pole-to-equator temperature difference of 27 K, the tropical maximum being 300 K and the extratropical value being 273 K. We consider the impact of including SGW parameterizations on this ocean simulation. We then perturb the tropical maximum and study the impact of waves across a range of different climate states. The notation for the experiments used in this paper reflects the Tropical Maximum SST (e.g. TM300). We choose these profiles since they elicit a strong response in the large-scale atmospheric circulation that mimics the geometrically simplest climate oscillation: annular modes. This oscillation is reflected in a weakening and shift in the jet streams and expanded Hadley circulation. Thus, they provide an idealised framework for studying wave-induced mixing responses to climate variability.

2.2 Surface Gravity Wave Contributions to the Ocean Surface Boundary Layer

The temporal evolution of TKE (symbolically *e*) is the result of production through vertical shear and SGWs, destruction/production through stratification, vertical diffusion, and dissipation:

$$\frac{\partial e}{\partial t} = D_e + S + B + W - \varepsilon. \tag{5}$$

The right-hand side represents turbulent and viscous transport $D_e$, production by shear $S$, buoyancy $B$, SGWs $W$, and dissipation $\varepsilon$ represented as follows:

$$D_e = \frac{\partial}{\partial z} \frac{v_t}{\sigma_e} \frac{\partial e}{\partial z}, \tag{6}$$

$$S = v_t \left[ \left(\frac{\partial u}{\partial z}\right)^2 + \left(\frac{\partial v}{\partial z}\right)^2 \right], \tag{7}$$

$$B = -K_t N^2, \tag{8}$$

$$W(z) = \frac{w_{LC}^3}{H_{LC}}, \tag{9}$$

$$\varepsilon = \left(C_\mu^0\right)^3 \frac{e^{3/2}}{l}, \tag{10}$$

where $\sigma_e$ is a constant Schmidt number (the ratio between momentum diffusivity and mass diffusivity), $v_t$ and $K_t$ – vertical turbulent viscosity and diffusivity, $N$ – Brunt–Väisälä frequency, $l$ – mixing length and $C_\mu^0$ is a constant. The origin and derivation of the non-SGW expressions is relatively standard and can be found in Madec et al. (2016).

The SGW source term $W$ for TKE represents Langmuir circulation parameterized following D'Alessio et al. (1998) and Axell (2002) using the vertical velocity profile for Langmuir cells (LC) defined as:

$$w_{LC} = \begin{cases} c_{LC} u_s \sin\left(-\frac{\pi z}{H_{LC}}\right); \text{when } H_{LC} \geq -z \\ 0; \text{otherwise} \end{cases} \quad (11)$$

where $c_{LC}$ has a constant value of 0.15 (Axell, 2002), $u_s$ is the Stokes drift, and $H_{LC}$ is the LC depth. $H_{LC}$ is defined as the depth to which a water parcel with kinetic energy due to Stokes drift can reach on its own by converting its kinetic energy to potential energy, according to

$$-\int_{-H_{LC}}^{0} N^2 z \, dz = \frac{1}{2} u_s^2. \quad (12)$$

An additional effect of SGWs on $\partial e/\partial t$ is the energy flux from breaking waves. This is introduced as a Dirichlet boundary condition on TKE (Mellor and Blumberg, 2004). Furthermore, we account for the Stokes-Coriolis force – the interaction between wave-induced Stokes drift and the Coriolis effect (Hasselmann, 1970). This is introduced as an additional term in the momentum equation. This force alters the current profile throughout the entire Ekman layer (Polton et al., 2005).

In our OGCM setup, wind wave equilibrium is not assumed, i.e. the waterside stress is not necessarily equal to the total atmospheric stress. Instead, the ocean is mechanically driven by the difference between the total atmospheric stress and the net stress going into the waves. To account for this in the OGCM simulations with SGW effects included, the momentum flux computed by NEMO from the AGCM output ($\tau_{total}$, Eq. 1) is scaled by the ratio $\tau_{air-side}/\tau_{water-side}$. This ratio computed manually from SWM output and then passed as input to NEMO. WW3 computes $\tau_{air-side}$ and $\tau_{water-side}$ for each wave frequency individually. $\tau_{air-side} = u_*^2$ where $u_*$ is the friction velocity attributed to air-side stress computed from 10 m winds using an empirical formula described by Wu (1982). $\tau_{water-side}$ is the SGW-to-ocean momentum flux defined within the SWM wave dissipation term (WW3DG, 2016). In the wave model, the relationship between air and water density is assumed constant so is dropped from the air-side to water-side ratio.

We do not account for the sea-state dependent TKE flux used by Breivik et al. (2015) as well as the Stokes drift impact on the mass and tracer advection (Wu et al., 2019) since these have not been implemented in the latest NEMO release. We also do not account for non-breaking wave induced mixing (Qiao et al., 2004) since its current implementation in NEMO has not been widely validated against observations (see Appendix A).

The experiments which include and do not include the SGW parameterizations in the OGCM are referred to as '*waves*' and '*no waves*' respectively.

**3. Control Simulations**

3.1 Steady and Transient Aspects of the Control Climate

We begin by examining the control experiment designed to correspond to Earth's contemporary Northern Hemisphere (NH) summer (TM300, Figure 1). The conceptual interpretation of this control climate, simulated without SGW parameterization, underpins our latter examination of SGW influences (Section 3.2) and behaviours in cooler and warmer climates (Section 4).

NH summertime conditions are manifest in the warmer temperatures there, the weak midlatitude atmospheric jet stream (centered on 40 °N, Figure 1e) and the protrusion of the strong Southern Hemisphere (SH) Hadley cell to 10 °N (Figure 1f). Evaporation minus precipitation (E-P) is negative in the Intertropical Convergence Zone (ITCZ) region at 10 °N (Figure 1b) and is positive in both hemispheres' subtropics. The ocean is freshest (~30 PSU) around ITCZ latitudes, inline with the E-P minima marking positive freshwater fluxes from tropical precipitation (Figure 1d). This global salinity minimum is co-located with the global ocean's highest potential temperature and the area of very strong oceanic stratification (explicitly shown in Figure 6a).

The midlatitude NH upper ocean (down to 1000 m) is generally warm (~295 K) and salty (~35 PSU), while the SH ocean is cooler (~280 K) and comparatively fresh (~32 PSU). SH winter temperature inversion is identified southward of 40 °S. MLD is largest in the SH particularly in subpolar latitudes following the large wintertime heat loss there (see Figure 9d).

The thermocline depth in the tropical oceans is marked by sharp potential temperature decrease around 1000 m and with a corresponding relatively strong salinity gradient at that depth in the SH. The salinity gradient around 1000 m in the NH (i.e. summer) is much weaker. In the SH subtropics (5-30 °S), local salinity values are low (~32 PSU) in comparison to the NH subtropics (~35 PSU) (Figure 1d). However local E-P values in the SH subtropics are ~x3 the magnitude of the E-P values in the NH subtropics (Figure 1b). This indicates that the ocean is effectively advecting salinity away from the surface ocean in the SH subtropics.

The tropical ocean density structure in the control experiment is characterized by equatorial upwelling of relatively cold water extending almost to the surface (~10 m). This well-mixed cold water is associated with meridional overturning circulation (MOC) transporting water up from the abyss to circulate southward at the surface (Figure 1h). We interpret this tropical surface circulation pattern as southward Ekman transport forced by the easterly trade winds (Figure 1e). This southward Ekman flow converges with the streamfunction changing sign around 25 °S (Figure 1h) triggering downwelling and thus marking the southward edge of the strong MOC cell and defining the bounds of the warm tropical ocean (Figure 1c).

This midlatitude equatorward Ekman flow is likely forced by the strong SH midlatitude jet stream westerlies projecting on to surface momentum flux (Figure 1e). The mass convergence of the two Ekman flows terminates the strong wind-driven subtropical cell characterized by maximal mass transport of 104 Sv (the real Earth's equatorial current is 40-120 Sv in the Pacific Ocean; Tabata, 1975). Another anomaly of colder and saltier water identified on the northward edge of the ITCZ (Figure 1c,d) is presumably driven by similar Ekman dynamics to its stronger, SH counterpart. The strengths of the major oceanic overturning cells from south to north in Figure 1h are: 41, -104, 15 and -26 Sv. Despite the absence of a thermohaline circulation extending from the tropics to high latitudes, the aquaplanet ocean is efficient in transporting heat from the tropical to subpolar latitudes via Ekman transport consistent with findings of Marshall et al. (2007).

What is the transient evolution of zonal winds associated with this climate state? The temporal evolution of surface zonal wind is particularly relevant for establishing large magnitude, coherent packets of SGWs (Sverdrup and Munk, 1947). In the midlatitudes the surface westerlies in both hemispheres display very different transient characteristics associated with strong seasonality in midlatitude baroclinicity and atmospheric planetary wave activity (Figure 2b,g). Supplementary Figure S3 shows the probability density functions, means ($\mu$) and standard deviations ($\sigma$) of these surface winds. The SH midlatitude winter mean flow is 4 ms$^{-1}$ and very variable ($\sigma$ = 6 ms$^{-1}$) with the zonal wind direction there often changing sign (Figure 2b). This follows from the strong eddy activity in the storm track region. This inhibits the generation of a midlatitude Ekman flow of comparable magnitude to that seen in the tropics. Meanwhile, the NH's summer westerlies (Figure 2g) are stronger in the time-mean and steadier ($\mu$ = 6 ms$^{-1}$, $\sigma$ = 3 ms$^{-1}$).

The strong SH easterlies ($\mu$ = -7 ms$^{-1}$, Figure 2c) show signals of planetary waves that have their origin in the midlatitude westerlies belt (Figure 2b) however SH trade winds are generally steady ($\sigma$ = 3 ms$^{-1}$). The strong, steady nature of these trades constitutes the principal driver of the strong SH subtropical oceanic MOC (Figure 1h). The weaker meridional temperature gradient and (related) comparative absence of coherent planetary wave structure in the summer hemisphere results in weaker NH easterlies ($\mu$= -4 ms$^{-1}$, Figure 2f).

The ITCZ region (5-15 °N; Figure 2e) shows westward propagating mesoscale activity (-10 to 10 ms$^{-1}$). The structure here is reminiscent of convectively-coupled equatorial waves and inspection of precipitation's frequency-zonal wavenumber space shows high spectral density for equatorial Rossby waves. Despite this highly transient behaviour, the ITCZ mean flow is effectively still. The weaker NH jet (Figure 2g), trades (Figure 2f) and the ITCZ's highly variable winds (Figure 2e) result in a much weaker NH Ekman flow and corresponding tropical oceanic overturning structure (Figure 1h).

What do the SGWs corresponding to this momentum flux field look like? Recall that these waves are not parameterized into the control experiment's ocean simulation discussed above. The mean significant wave height field, $H_s = 4\sqrt{E}$, $E$ being the integral over the wave energy spectra, exhibits the largest amplitudes in the extratropics (~4 m) and the smallest in the deep tropics (~2 m) (Figure 3).

Wave heights are most variable and have globally maximal transient amplitudes (~8 m) in regions of midlatitude atmospheric baroclinicity, not where the mean winds are strongest (Figure 3b,g; Supplementary Figure S4). While the SH trades are 50% stronger than the mean SH midlatitude surface winds, mean $H_s$ in the SH subtropics is lower (3 m vs 4 m). This is understood by noting that although the atmospheric baroclinic eddies originating in midlatitude storm track cause $H_s$ to be twice as 'choppy' there - σ is 1.4 m vs 0.7 m in the subtropics - these eddies propagate synchronously in the direction of the mean flow. Conversely, in the subtropics they act against the mean flow, inhibiting SGW growth. It is possible to identify imprints of some of the strongest tropical atmospheric disturbances in $H_s$ for example in Figure 3e around time 0002-05, 100 °lon and 3-01, 50 °lon.

The temporal evolution of the surface Stokes drift qualitatively manifests much of the same apparent variability as the surface winds (Figure 4 vs Figure 2). While the impact of midlatitude baroclinic eddies is more muted in SH subtropical $H_s$ field (Figure 3c) it is very clear in the corresponding Stokes drift field (Figure 4c).

3.2 Impact of Surface Waves on the Control Oceanic Mean State

The ocean's response to SGW parameterization is reflected in both its dynamic and thermodynamic state (Figure 5). We now consider the reference climate state (TM300) ocean simulations with and without SGWs.

At the surface, the inclusion of SGW parameterization reduces momentum fluxes everywhere (for more details see Section 2.2). This slows down the climatological zonal currents by 2-4 % (Figure 5b). These values are in agreement with other studies which find this effect reaching up to 5 % (Janssen et al., 2013) and even up to 20 % in winter (Gulev and Hasse, 1998) in the open tropical ocean. Away from the deep tropics, this dampening effect penetrates to 100 m. In the deep tropics however, the impact of dampened surface momentum fluxes extends to almost 1000 m but the structure here is not simply slowing the current everywhere. Recall the dominant zonal feature is the strong easterly current extending to ~2000 m depth across the equator between 10 °S and 10 °N associated with the SH Hadley cell protrusion into the NH. The northern component of the current becomes rapidly weaker below 50 m, while the southern component's strongest current velocity penetrates to ~1000 m. With the surface momentum flux dampening by the SGW parameterization, the northern component of this current monotonically dampens at all depths to 1000 m. Between 10 °S and the equator, where the climatological current is strongest (~1 ms$^{-1}$), the current is dampened by ~1 % down to 10 m, below which the dampened surface momentum flux results in increasing the zonal current velocity by ~1 %.

SGWs amplify existing vertical momentum diffusivity generally between 20 and 100 m deep (Figure 5d). The strongest impact of waves on vertical momentum diffusivity is in the NH high latitudes (+70 °N) where the ocean is vertically homogenous in terms of temperature and salinity in the upper 1000 m in both '*waves*' and '*no waves*' experiments (Figure 5e,f,g,h). The other areas with strong SGW impact are the midlatitudes of both hemispheres (30-50 degrees latitude) where they increase it by 30-50 %. A large but smaller contribution of SGWs (~20 %) is found around 10 °S. These contributions are co-located with local maxima in momentum flux, surface Stokes drift and mean wave heights (see Figure 9).

These mechanical impacts markedly alter the temperature, salinity and stratification structure of the ocean down to ~500 m (Figure 5f,h; Figure 6b). Waves deepen the pycnocline in the midlatitudes and SH subtropics and reduce stratification in the upper 200 m (Figure 6b). These stratification changes result from both temperature and salinity alterations although in percentage terms salinity changes (~1 %) are an order of magnitude larger than temperature changes (~0.1 %).

Between 50-30 °S, the weakened zonal current and increased vertical momentum diffusivity results in negative temperature and salinity anomalies (Figure 5f,h) tracing the climatological halocline (Figure 5g) between 100 and 200 m. This is associated with increased salinity (Figure 5h) and reduced stability (Figure 6b) in the midlatitude upper ocean strongly aligned with downwelling at the convergence of the tropical and subtropical Ekman driven MOC cells (~25 °S, Figure 1h). This mirrors changes in the subtropics seemingly following from reduced upwelling and southward transport of warm equatorial waters: deepening of the thermocline and cooling of the surface waters around 5 °S and reduced salinity (recall salinity in the subtropics is imported at depth from the midlatitudes by MOC). At the ITCZ latitude, SGW parameterization results in a strengthening of both halocline and thermocline, impacting the already extremely strong stratification there by ~5 % (Figure 6b). Away from the ITCZ, the wave parameterization generally acts to weaken the upper ocean (>100 m) stratification by ~30 %.

Subtropical and extratropical oceanic vertical shear is reduced by ~50 % in the upper 100 m, but is very low there regardless (Figure 6c,d). Near the equator, where shear is orders of magnitude stronger, SGWs also reduce shear in the surface layer, and they weaken and broaden the finger of high shear values tracing the SH equatorial thermocline inside the strong Ekman MOC cell (Figure 6d).

SGWs may also induce stability changes through relative shifts in stratification and shear (Li et al. 2012). Here we consider differences in time-mean gradient Richardson number, $Ri = (N/S)^2$, with the key heuristic being that instability is associated with high shear and impeded by stratification. In general, shear-related instabilities are thought to occur where $Ri$ < 0.25 (Mack and Schoeberlein, 2004). This is true for the upper 40 m layer of the winter hemisphere (SH, Figure 6e). SGW parameterization has the effect of shoaling the maximum depth of the critical $Ri$ values in the SH extratropics (Figure 6f). It also weakly lowers $Ri$ in the SH subtropics. We note though that the ocean is partially covered with ice south of 48°S, so there is no local Langmuir parameterisation there. Just equatorward of this region the largest wave heights on the planet are observed (see Fig. 9c) and wave contribution to MLD is ~20% (see Fig. 9e). This shoaling could therefore be an impact of remote SGW effects. The current model, lacking SGW-ice coupling and having limited vertical resolution in the upper 50m, is not sufficient to explore this more.

3.3 Summary Understanding of the Control Climate

Overall, we find that our solstitial aquaplanet ocean is dominated by Ekman driven MOC cells associated with equatorial upwelling and midlatitude downwelling. The meridional extent of these cells is controlled by the strength and transient evolution of the atmospheric jet streams and Hadley circulation. The difference between the two statistically steady state climates with and without SGW parameterization shows that SGWs act to dampen Ekman driven MOC enough to change the density of the upper ocean with the strongest impact being on salinity, which changes by $O(1 \%)$, while temperature modulation by SGW effects are smaller being $O(0.1 \%)$. This reduces upper ocean stratification in regions where it is generally climatologically stable. This is associated with a deepening of the pycnocline in midlatitudes of both hemispheres.

## 4. Impact of ocean surface waves across different climates

We now consider the contribution of SGWs to the aquaplanet ocean across different climates. We change the simulated climate by varying the tropical SST maximum (Figure 7a) and thus the large-scale meridional thermal gradient. These climate variations are in some generic sense analogous to the Southern Annular Mode (SAM), the canonical metric for which is the difference in sea-level pressure between 40 and 65 °S (Gong and Wang, 1999). As the climate warms, the surface pressure gradient between these latitudes increases in both the winter and summer hemispheres (Figure 7c). The corresponding 'SAM' index values recovered in our simulations are shown in Supplementary Figure S5.

4.1 Colder and Warmer Solstitial Aquaplanet Climates

As tropical SST increases, the tropospheric jet stream intensifies and shifts poleward in the winter hemisphere (SH, 10 ms$^{-1}$ zonal-mean zonal wind velocities are shown in contours in Figure 7b). The magnitude of these shifts can reach up to 10 degrees latitude in particular simulations. In the summer hemisphere (NH), the jet stream intensifies without shifting, except under the most extreme tropical SST perturbation. In the wintertime SH, poleward jet shifts under warmer climates correspond to Hadley cell expansion (both from the southern terminus moving southward and the ITCZ moving northward) and poleward shifts in the polar cell (Figure 7d). In the summer hemisphere (NH), this direct jet stream - Hadley extent association is also visible, being however less evident than in the SH. In line with this, the "wet-gets-wetter, dry-gets-drier" (WWDD, Byrne and O'Gorman, 2015) pattern is seen in E-P (Figure 7e).

These changes to the zonal-mean aquaplanet circulation across climate warming, specifically the jet stream shifts and Hadley cell expansion, are analogous to the real Earth case (Staten et al., 2018, Shaw, 2019). They are therefore useful simplified simulations for understanding changes in SGWs under large-scale climate change. The poleward jet and storm track shifts are well understood (for discussion of the mechanics see Chemke and Kaspi, 2015; Tamarin and Kaspi, 2017; Shaw, 2019), while the mechanics of tropical expansion is less clear (Staten et al., 2018).

As discussed in Section 3, the aquaplanet ocean is dominated by the position and strength of the jet streams and trade winds. With climate warming, the oceanic subtropical zonal jet becomes weaker and the SH midlatitude jet becomes stronger (Figure 8a, Supplementary Figure S6). This follows from both the increase and the poleward migration of wind stress maxima (Figure 9a). The extension of the strongest oceanic overturning cell in the SH subtropics becomes larger under warmer climates due to the poleward shift of the southern boundary of this cell tracing the jet shift (Figure 7f). The intensity of this cell weakens with both cooling and warming from the control experiment (Figure 8b). The positive overturning cell in the SH midlatitudes also weakens with warming from the control (Figure 8b) associated with poleward shifts in surface wind stress (Figure 9a).

MLD in the tropics and summertime NH midlatitudes decreases from 80 m to 40 m with warming from TM290 to TM310 (Figure 9d). The moderate MLD amplification with climate warming in the SH extratropics can be understood through an increase in surface wind stress (Figure 9a), however, while wind stress changes by a factor of 2 in simulations with different SST gradients, MLD changes only by a few percent. Meridional shifts in SH $H_s$ maxima (Figure 9c) are consistent with the surface stress (Figure 9a), however the wave response to the changes in large-scale atmospheric circulation is non-linear, especially in equatorial and subpolar latitudes.

When integrated over the mixed layer, global mean buoyancy frequency $N^2$ – i.e. upper-ocean stratification - increases with climate warming (Figure 8c). This is consistent with recent observational studies of global warming (Li et al., 2020). Locally, $N^2$ changes vary. In the summertime NH midlatitude jet region it increases logarithmically with warming, and the impact of SGWs on reducing $N^2$ also becomes larger under warmer climates. Global-mean mixed layer vertical momentum diffusivity is approximately constant with climate (Figure 8e). Locally however it is sensitive to climate, again consistent with observations (Busecke and Abernathey, 2019). In the SH (winter) it increases with warming while in the NH (summer) in decreases (Figure 8f).

4.2 Surface Wave Impacts across Different Climate States

Including SGW parameterizations into the aquaplanet OGCM impacts the oceanic climate selectively. Specifically, SGWs dampen the oceanic midlatitude zonal jets in all simulated climates by a few percent (Figure 8a) and dampen meridional overturning in the ocean surface boundary layer (Figure 8b, Supplementary Figure S7). SGWs increase global-mean vertical momentum diffusivity integrated over the mixed layer by ~40 % in all climates (Figure 8e). However, they only have an extremely subtle effect on global mean buoyancy frequency (Figure 8c). They do have a variable local impact on stratification: they increase $N^2$ integrated over the mixed layer in the SH midlatitudes (Figure 8d), but decrease it in the NH hemisphere midlatitudes. The effects of SGWs on local mixed layer integrated buoyancy frequency and vertical momentum diffusivity appear variable with climate.

In simulations with SGWs, MLD has larger magnitudes almost everywhere except at the equator (Figure 9e) i.e. where upwelling occurs (Figure 1c). SGW contribution to MLD has a peak around 45 °S – i.e. exactly where wave heights are largest (Figure 9c). This is likely associated with the wintertime atmospheric jet stream (Figure 7b). In simulations with SGWs, the mixed layer in the SH midlatitudes is up to 20 % deeper than in the '*no waves*' simulations. Meanwhile MLD in the NH is deepened by up to 10 % by the SGWs.

Attempts to derive scaling laws for MLD controls are ongoing and the role of SGWs is actively debated (e.g. Pearson et al., 2018). We have binned zonal-mean diagnostics into different thermal and rotational regimes defined in Figure 2. We find a strong linear relationship ($R^2$=88 %, $p$<<0.05) between climatological MLD and the simulated contribution of SGWs to MLD. Across all our simulations and climate states, SGW parameterization generally adds ~10 % to MLD (Figure 10a). This is most simply related to surface Stokes drift and friction velocity (Figure 10c,d).

We find that across the simulated climate states, the SGW parameterizations generally add an additional 3 m to MLD when significant wave heights increase by 1 meter (Figure 10b). However, the spread around this climate-agnostic linear relationship is large ($R^2$=42 %, $p$<<0.05). This spread is partly explained by differences across climates (represented by colors in Figure 10). In the warmer climates $H_s$ values increase, but for the lower magnitude $H_s$ values (2 to 3 m) the addition to MLD by SGW parameterization decreases with larger values of $H_s$ - i.e. the opposite to the general relationship (Figure 10b). In some cases, MLD is larger without SGW parameterization than with. Even so, when considering each climate separately, there is a strong positive correlation that matches the climate-agnostic correlation.

Although globally surface Stokes drift and friction velocity are positively correlated to SGW parameterization increasing the MLD ($r$=0.69 and 0.71 respectively, $p$<<0.05 for both), locally this is not necessarily true (Fig. 10c,d). For example, between 50 and 65 °N (i.e. the NH storm track region), climate warming increases both surface Stokes drift and friction velocity but the SGW parameterization contribution to MLD declines with climate warming. So overall we find that on the aquaplanet, climate changes can impact the local role that SGWs play in MLD in a way that overwhelms a climate-invariant relationship with SGWs and MLD.

**5. Summary and Discussion**

5.1 Synthesis

We examine how the inclusion of SGW parameterizations impact solstitial aquaplanet oceans in terms of their mean thermodynamic and dynamic state. In a control simulation analogous to modern-day Earth we find that the ocean circulation is dominated by an equatorial zonal jet and four Ekman driven overturning cells aligned with the two westerly jet streams and easterly trade winds in both hemispheres. The equatorial upwelling and midlatitude downwelling associated with this Ekman forcing brings cold water up at the equator and forces warm surface waters down in the midlatitudes to around ~1000 m deep. There is no meaningful thermohaline circulation.

Including SGW parameterization increases vertical momentum diffusivity in the upper ocean by around 40 % and dampens the surface momentum fluxes by ~4 %. This represents a redistribution of upper ocean energy: lowering vertical shear and dampening the Ekman driven MOC cells but also pushing MLD down ~10 % and decreasing midlatitude stratification. The midlatitude pycnocline in both hemispheres is deepened by the inclusion of SGWs. The dampened overturning changes the density structure of the upper ocean by lessening the large-scale advective transports of heat and salt, thus freshening the equatorial latitudes (where E-P is negative) and providing more salinification in the subtropics (where E-P is positive) by changes $O(1 \%)$. Generally, these impacts are found to be consistent across cooler and warmer aquaplanet climates. In the control climate, relative changes to climatological shear and buoyancy frequency act to make the low Richardson number layer between 70 and 50 °S (winter hemisphere) shallower.

While the set of experiments presented and discussed here imply that SGWs play a non-negligible role in aquaplanet ocean dynamics, comparative assessment of different SGW parameterizations reveals consequential diversity in impacts in even the simplest idealized setups (Li et al., 2019) as well as more complex models (Morim et al. 2019). Thus, these results should not be considered conclusive.

The zonally symmetric idealisation of the aquaplanet has provided important advances in understanding regional climate dynamics (Collins et al., 2018; Maher et al. 2019). However, aquaplanet climates are extremely sensitive to model resolution and parameterization of atmospheric convection (Retsch et al., 2019). We have used a non-coupled ocean modelling approach which tends to simulate warmer and saltier upper ocean states than coupled simulations (Tréguier et al., 2005; Rattan et al., 2010, Gavrikov et al., 2020). In spite of these limitations to the model configuration, we suggest that the most significant takeaways about the role of SGWs in aquaplanet oceanic climate are robust. This role includes dampening the Ekman circulation, increasing momentum diffusivity and increasing MLD.

5.2 Relevance to Significant Climate Questions

Given potential climate impacts of SGWs, parameterizing their effects into GCMs could provide significant improvement of current MLD biases and perhaps others. As CMIP6 model data begin to be analysed, a question regarding SGWs is presented: is the exclusion of surface wave processes implicated in the nagging 'problematic aspects' in modelling the global ocean [quotation from Held et al. (2019) in discussion about the limitations of GFDL's CM4.0 CMIP6 contribution model in reproducing Southern Ocean variability]. Since SGW heights are projected to increase in the future, particularly in the Southern Ocean (Fan et al., 2013; Hemer et al., 2013; Wang et al., 2014; Morim et al., 2019; Young and Ribal, 2019, Lemos et al., 2020), changes in wave impacts under different climate states are potentially important. Breivik et al. (2019) showed that Stokes drift is projected to increase in the Southern Ocean by about 15 % by the end of this century which implies changes in turbulent Langmuir number and potentially increased mixing.

As we note in Section 4, the climates we explore here resemble the SAM and the Ekman driven MOC cells that dominate that aquaplanet ocean are not dynamically dissimilar from the Antarctic Circumpolar Current. In the warmest aquaplanet simulations in this work, Hadley cells extended poleward associated with a poleward shift in subtropical ridge. Both Hadley circulation extent and the SAM have been demonstrated to be dominant factors controlling wind-forced SGW climate in the Southern Ocean (Fan et al., 2014). A number of studies report shifts towards a positive phase of SAM and intensification of westerlies in both in contemporary climate change (Thompson and Solomon, 2002; Marshall et al., 2003) and projections of future climate (Zheng et al., 2013). This is driving trends in Southern Ocean wave climate.

We note there is a strong influence of the stratosphere particularly on the polar troposphere that is well documented to interact with SAM (e.g. Lim et al., 2019). By modulating the tropical SST maxima in the present runs we are changing both tropopause depth and the moistening rate of the stratosphere by convection and these both act to cool the tropical stratosphere (Domeisen et al. 2019; Supplementary Figure S8). The indirect impact of stratospheric temperature changes (inverse to their tropospheric counterparts) are only crudely represented in our model since most of the vertical layers are simulating the lower troposphere. We do not consider further this stratospheric pathway but note that this provides an additional impact across climate states and needs to be kept in mind when interpreting the results.

SGW interactions with sea ice, particularly in the Arctic Ocean, are an important mechanism within large-scale climate variability and change on Earth (Squire 2020). Moreover, sea-ice dynamics have been shown to be of great significance for ocean-covered exoplanet climate (Yang et al., 2020). Unfortunately, the simulations presented do not have the sea-ice model dynamically coupled to the SGW model. The dramatic responses of aquaplanet SGWs to large-scale climate changes shown here and their impact upon the oceanic mean state demonstrate that SGWs could be significant for research into exoplanetary oceans.

**Appendix A: A note on wave-induced vertical kinematic diffusivity**

Wave-induced vertical kinematic diffusivity (Qiao et al., 2004) is used in some studies and attempts to capture mixing induced by non-breaking SGWs by enhancement of pre-existing turbulence. Like Langmuir mixing, this effect is also associated with non-breaking waves but is, at least conceptually, distinct from Langmuir turbulence. While the $B_v$ concept has been useful in improving the characteristics of vertical thermal structure in OGCMs (Qiao et al., 2004,2010), how to implement it in OGCMs as a parameterization is not straightforward. In particular, the parameterization of $B_v$ that is currently implemented in NEMO introduces a coefficient proportional to Stokes drift velocity, wave amplitude and wave number and adds this directly to the OGCM vertical viscosity and diffusivity coefficient. This implementation has not been widely validated against observations yet (see Supplementary Figure S9 showing that the effect from this parameterization leads to differences in temperature and salinity an order of magnitude larger than considered in our paper). Wu et al. (2015) demonstrated that simplistic formulations of $B_v$ lead to overestimation of this effect on mixing in the upper ocean resulting in larger biases in MLD relative to observational data. The best results are obtained when $B_v$ is computed from the full directional wave spectra, however, this is tricky to implement since it requires both modifications in SGW model source code which will allow it to provide $B_v$ as an output as well as modifications in OGCMs.

**Acknowledgments**

We are grateful to Stephen Griffies (Princeton University) and the two anonymous reviewers for their comments that helped to improve the manuscript. We thank Bernard Barnier (CNRS and Shirshov Institute of Oceanology, RAS) very much for extremely useful discussions and very generous help configuring NEMO for aquaplanet simulations. We also thank Oyvind Breivik (Norwegian Meteorological Institute), Fangli Qiao (First Institute of Oceanography, China), Alexander Babanin (University of Melbourne), Yulia Troitskaya (Institute of Applied Physics, RAS) and Andrey Glazunov (Institute of Numerical Mathematics, RAS) for other discussions regarding this project. All data produced and analysed in this study will be open access and available at final publication in the Dryad data repository. The code bases for the various models used here can be found at: Isca: https://github.com/ExeClim/Isca, NEMO: https://www.nemo-ocean.eu/ and WW3: https://github.com/NOAA-EMC/WW3. This work was funded by NASA (JS, 80NSSC21K0558), under the contract with Russian Ministry of Science and Higher Education #14.W0331.0006 (SG, analysis of ocean mixing processes) and by RSF grant 17-77-20112-P (MM).

FIGURES.

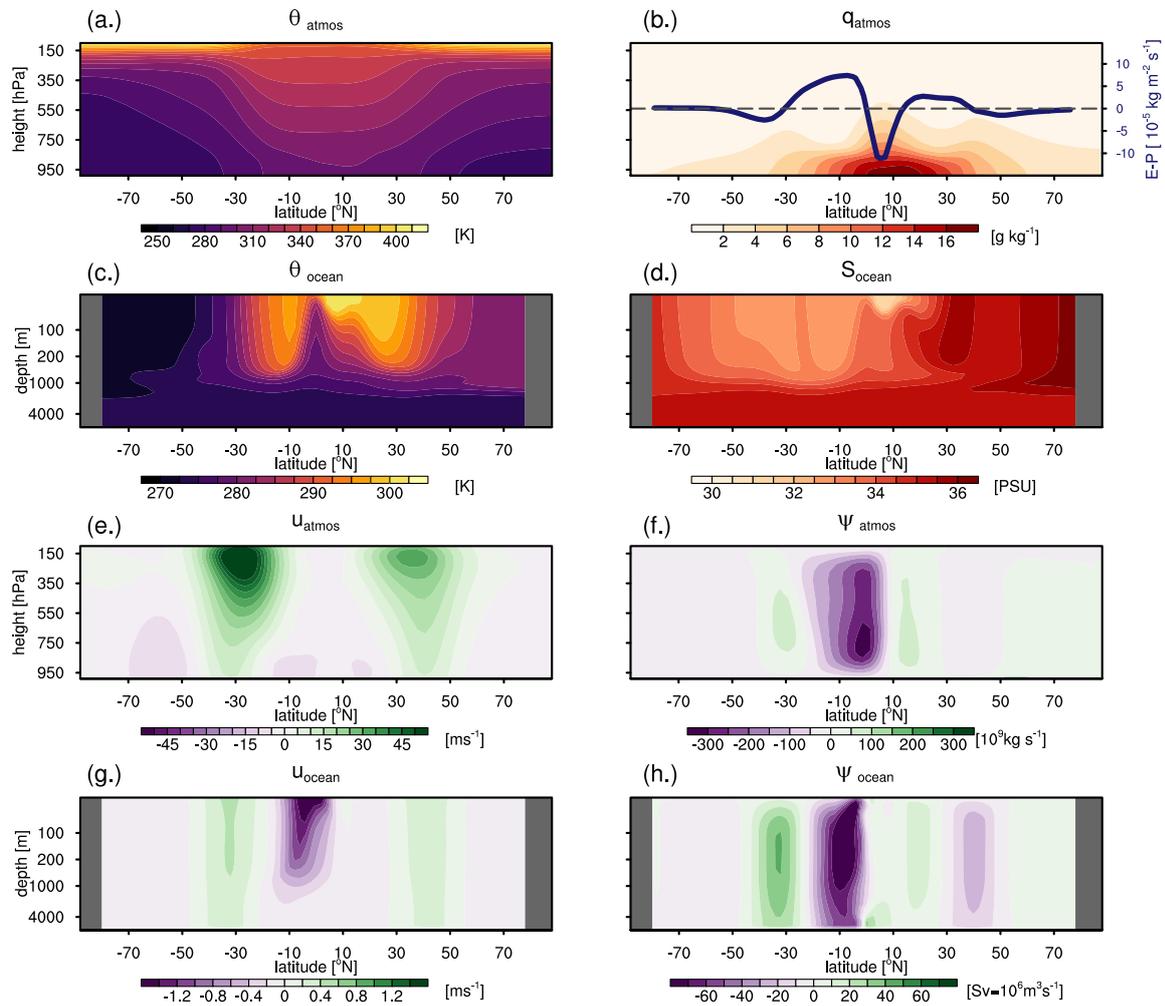

**Figure 1.** The TM300 control solstitial aquaplanet zonal-mean time-mean atmosphere and ocean. This simulation does not include SGW parameterization. Thermodynamic quantities: potential temperature in the atmosphere (a) and ocean (c); specific humidity and evaporation minus precipitation (b) and ocean salinity (d). Circulation measures: zonal wind in the atmosphere (e) and zonal current velocity in the ocean (g), meridional overturning circulation stream functions in the atmosphere (f) and in the ocean (computed using Eulerian velocities, h). Positive streamfunction values correspond to clockwise circulations. Bathymetry is shaded in grey.

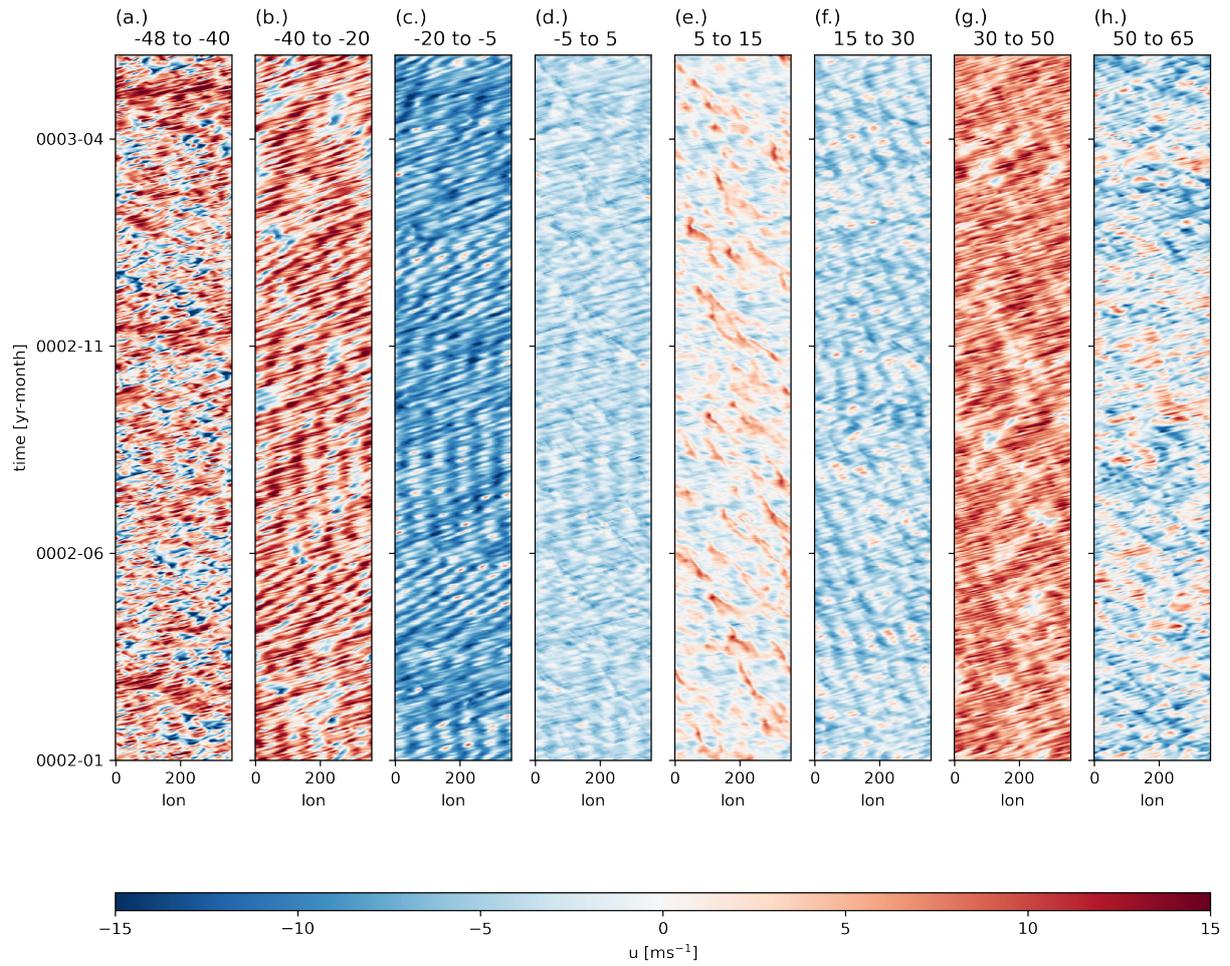

**Figure 2.** Hovmöller diagrams for meridionally-averaged surface zonal winds. Numbers in the subplot titles correspond to latitudinal bands over which the averages are computed. The surface is partially ice covered south of 48S (Supplementary Figure S2).

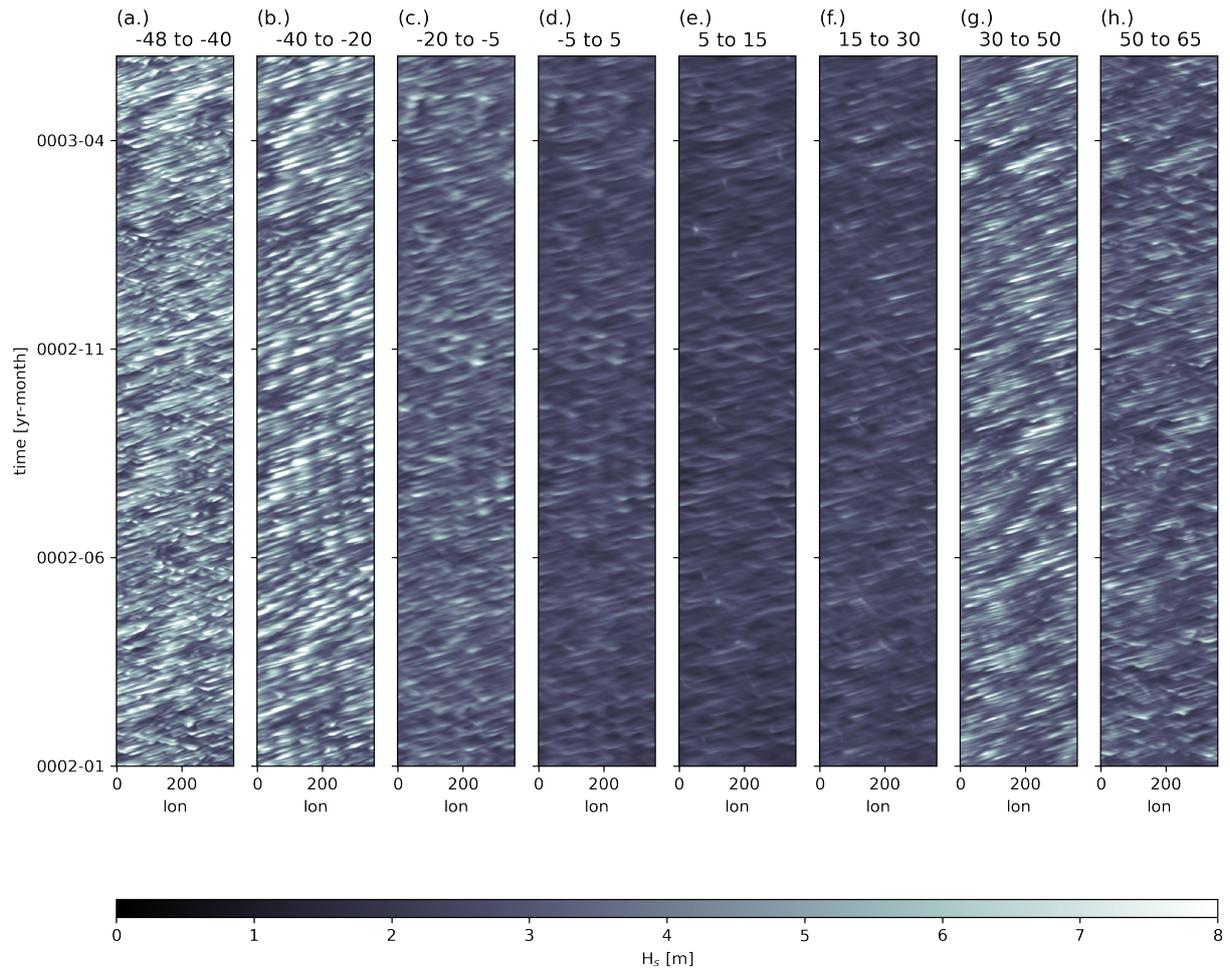

**Figure 3.** As in Figure 2 but for significant wave height, $H_s$.

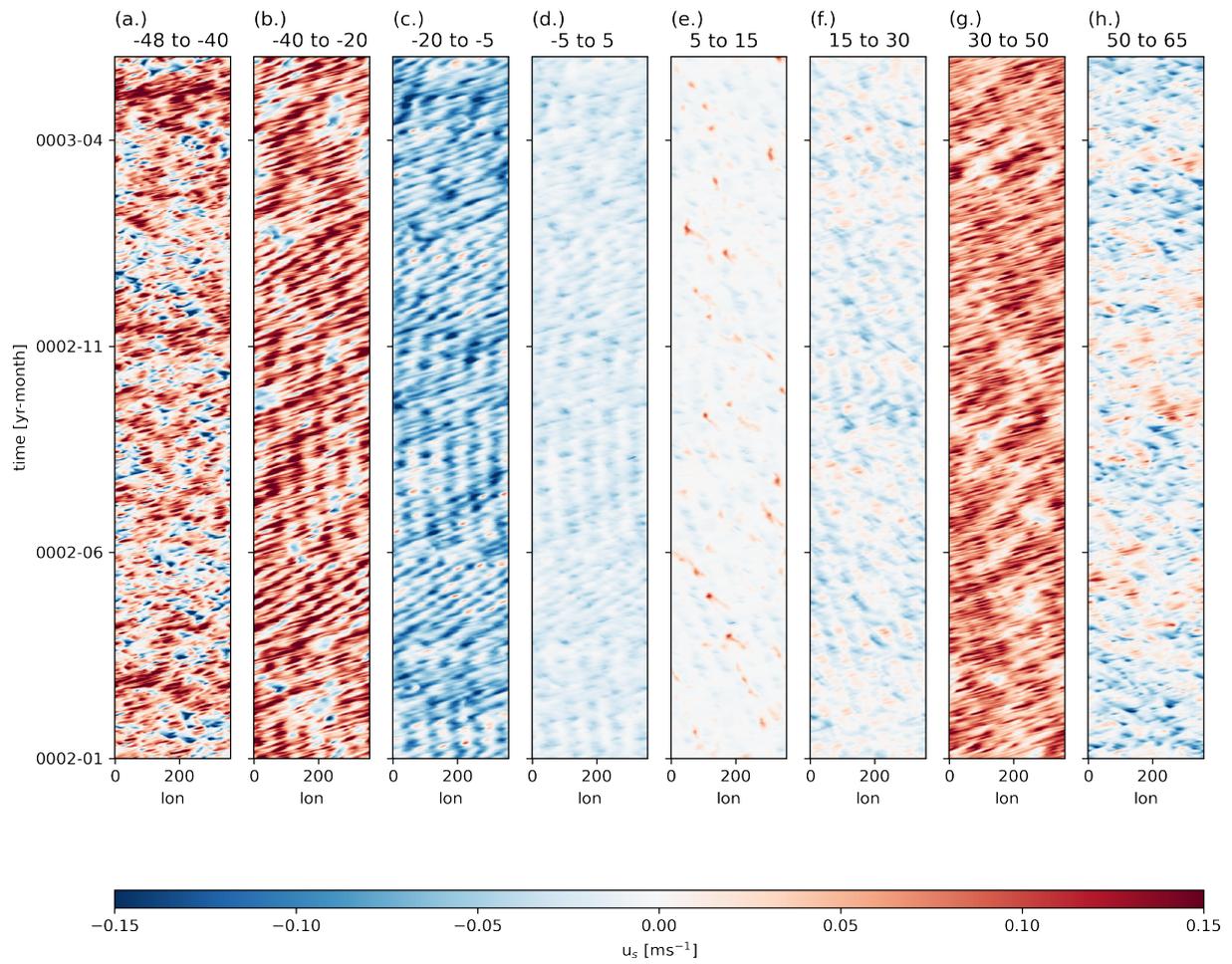

**Figure 4.** As in Figures 2 and 3 but for zonal Stokes drift at the ocean surface.

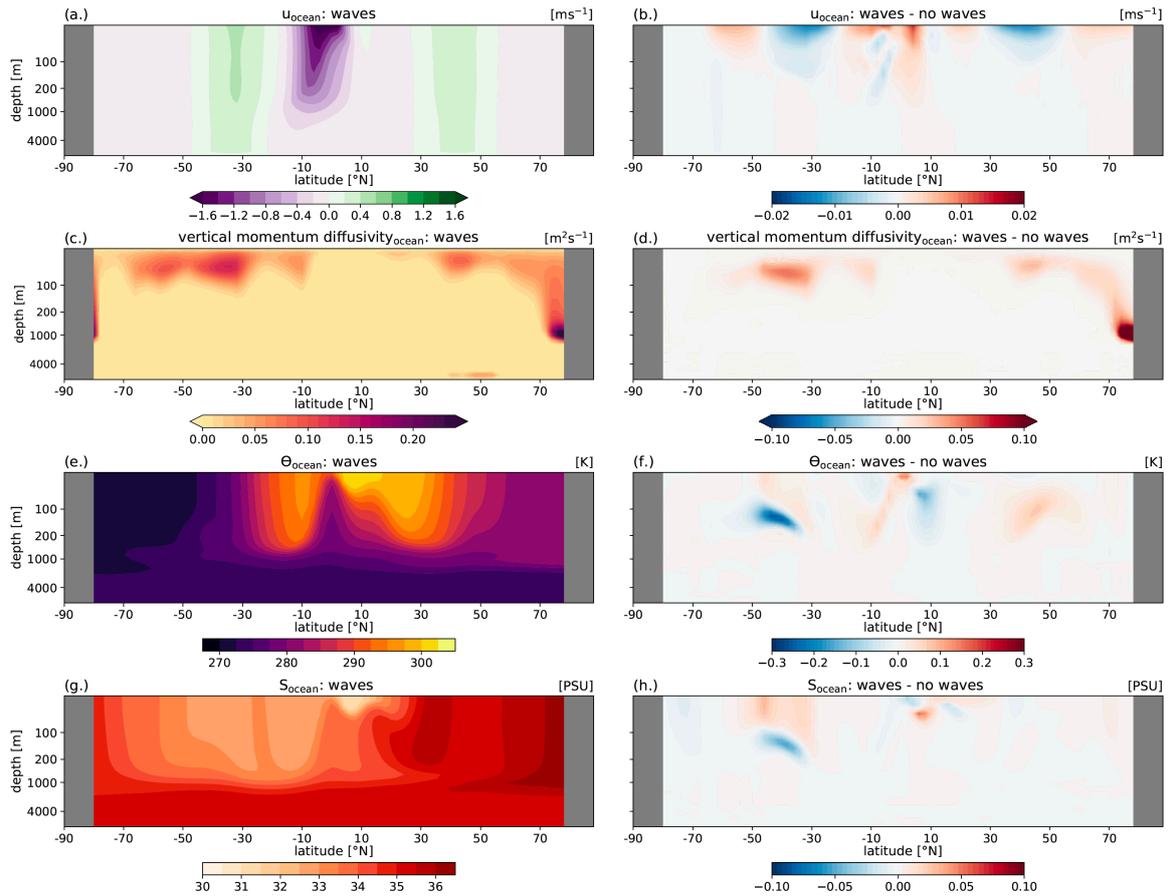

**Figure 5.** The TM300 aquaplanet ocean with SGW parameterizations. Zonal-mean zonal current velocity $u_{ocean}$ (a, b), vertical momentum diffusivity (c, d), potential temperature $\Theta_{ocean}$ (e, f) and salinity $S_{ocean}$ (g, f). The left column (a, c, e, g) corresponds to the climatological mean state. The right column (b, d, f, h) shows the differences between integrations with and without SGW parameterization.

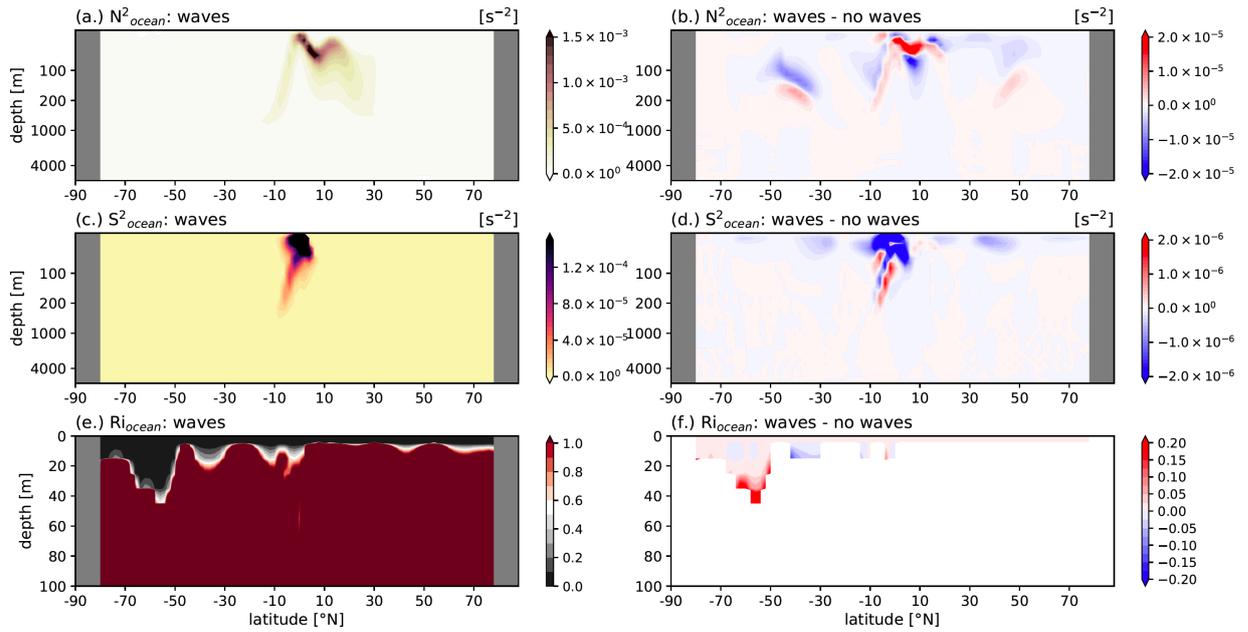

**Figure 6.** As in Figure 5 but for zonal-mean squared buoyancy frequency $N^2_{ocean}$ (a, b), squared vertical shear $S^2_{ocean}$ (c, d) and Richardson number $Ri_{ocean}$ (e, f). Note Ri is only shown for the upper 100m. In panel (f) we are masking regions where Ri>1 in *'waves'* simulation.

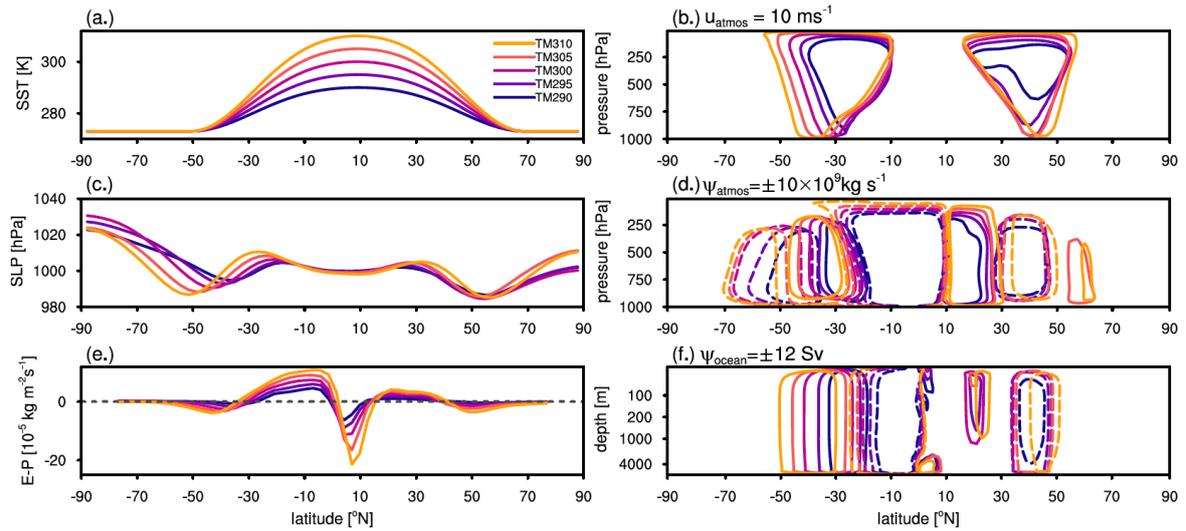

**Figure 7.** Simulated solstitial aquaplanets across different climates denoted by SST tropical maximum (TM, plotted in different colors). Left column shows surface quantities: (a) SST, (c) sea level pressure, and (e) evaporation minus precipitation. Right column shows particular contours (given in the plot titles) for (b) zonal mean zonal wind velocity, (d) the atmospheric meridional overturning streamfunction, and (f) the oceanic meridional overturning streamfunction (*'no waves'* simulations).

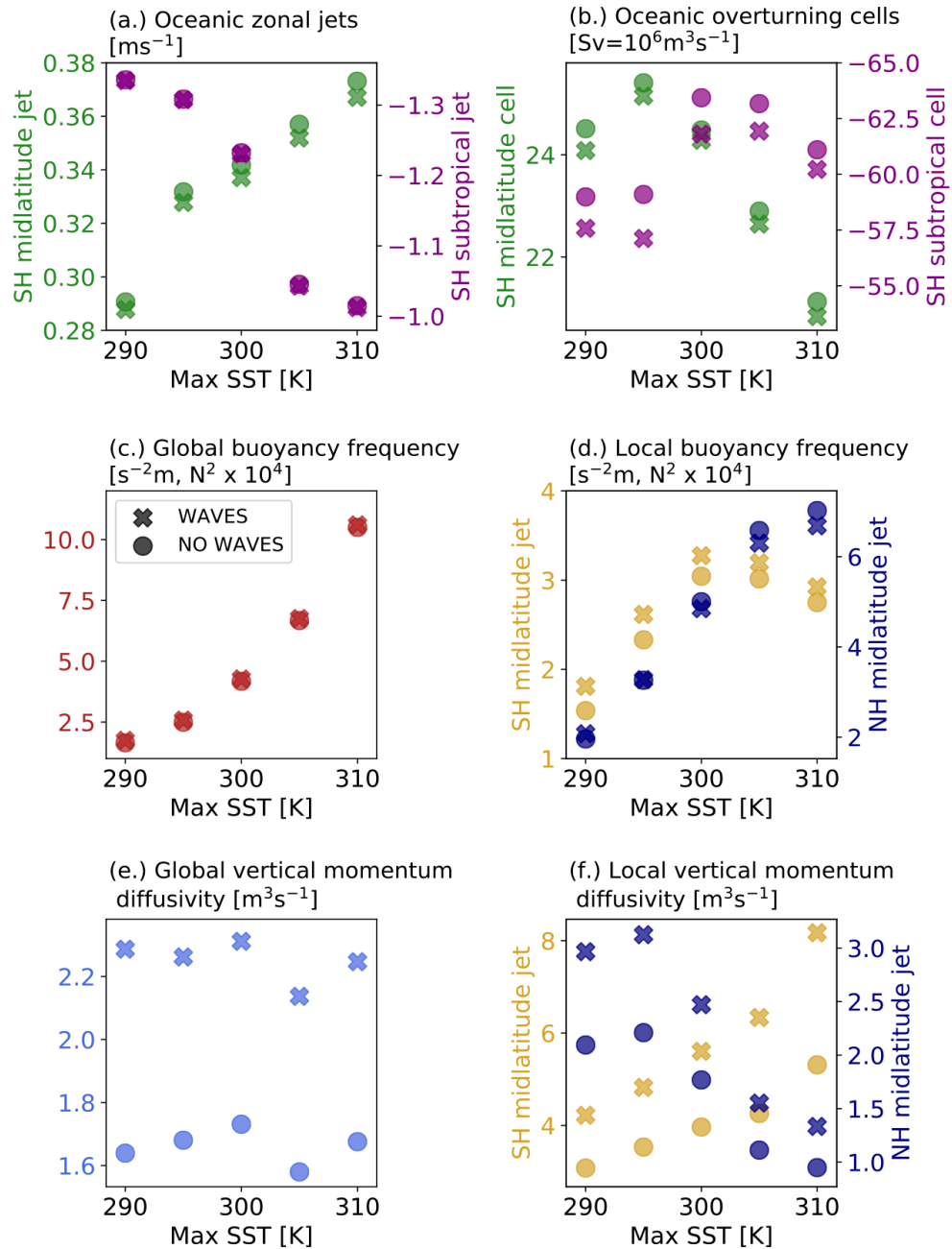

**Figure 8.** Diagnostics for near-surface oceanic zonal jet strength (a), near-surface meridional overturning strength (b), mixed layer buoyancy frequency (c,d) and mixed layer vertical momentum diffusivity (e, f) across different climates with (*crosses*) and without (*dots*) SGW parameterization. Each marker represents a diagnostic computed from the time-mean zonal-mean of a simulation. The oceanic jet and overturning cell boundaries are defined as the points where the zonal-mean zonal current velocity and the streamfunction respectively fall below 50% of the maximum (midlatitude) and minimum (subtropical) hemispheric values. See Supplementary Figures S6 and S7 for visualization of these jets and cells. The strength of the zonal jets and overturning cells are defined as the mean values of the respective quantities within the feature boundaries from 0 to 100 m. Buoyancy frequency and vertical momentum diffusivity have been integrated over the mixed layer defined following the turbocline depth criterion (Section 2.1.2). Global buoyancy frequency and vertical momentum diffusivity correspond to averages between 70°S and 70°N, local values are averaged over SH and NH midlatitude oceanic jets. Averages over latitude bands are cosine weighted.

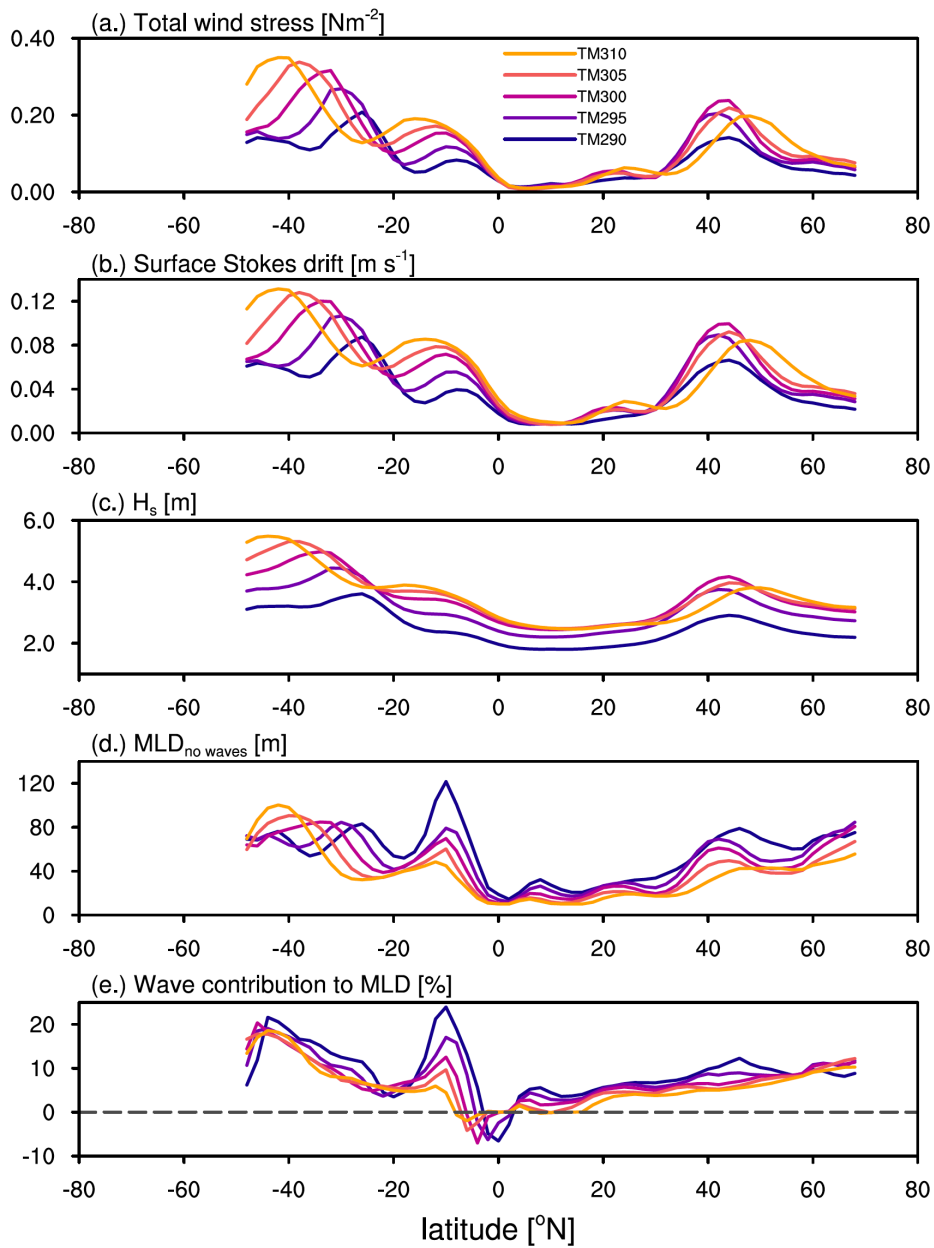

**Figure 9.** Total atmospheric wind stress (a), surface Stokes drift (b), significant wave height $H_s$, MLD in the '*no waves*' simulations (d) and SGW contribution to MLD across different climates (e) between 70 °S and 70 °N (signified by the different colors). The surface is partially ice covered south of 48S (Supplementary Figure S2).

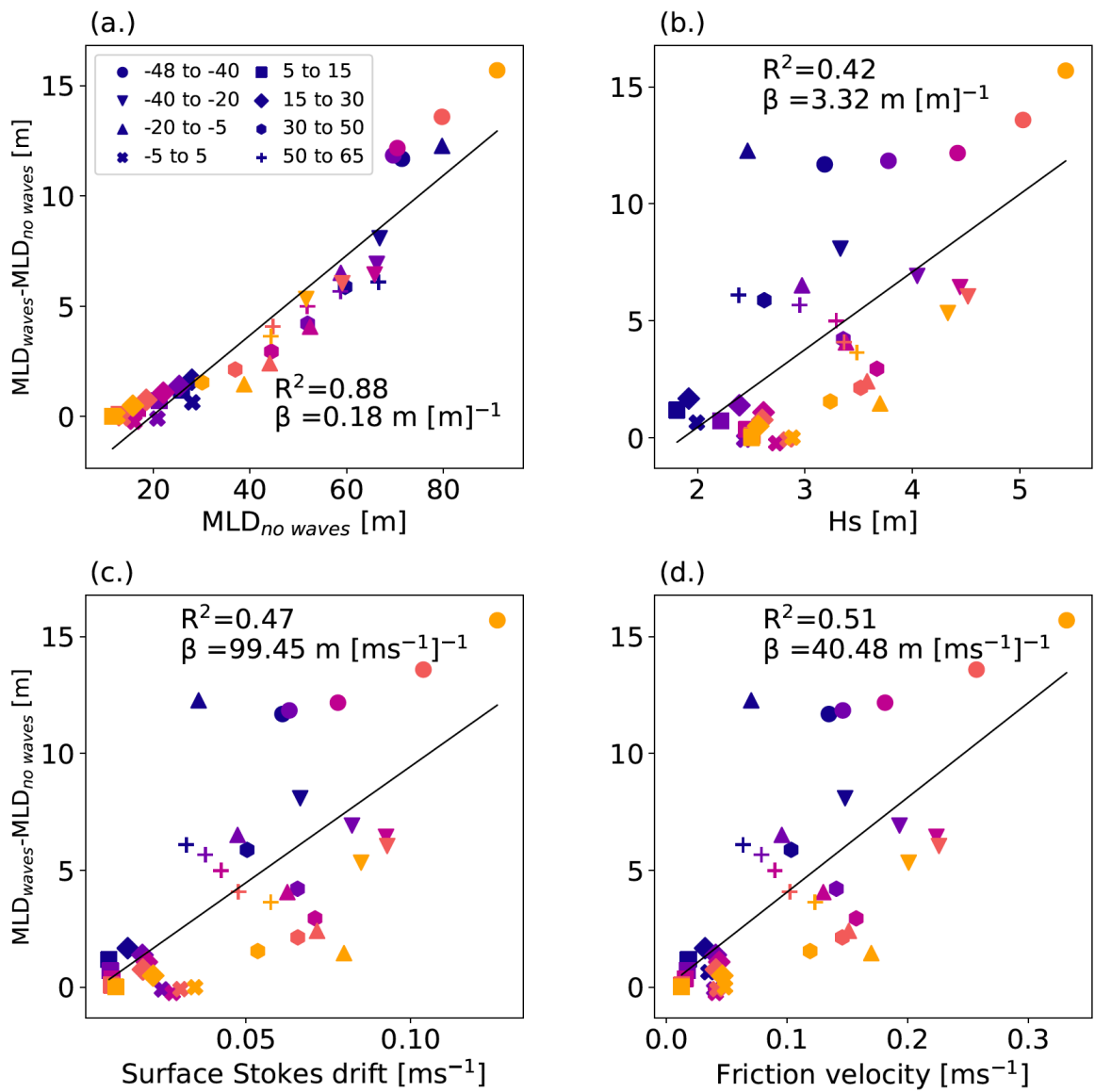

**Figure 10.** SGW contribution to MLD as the difference between MLD in the '*waves*' and '*no waves*' simulations. Colors correspond to the different climates as defined in Figure 7. Each point corresponds to a zonal and cosine-weighted meridional mean in different thermal and rotational regimes as divided up in Figures 2,3,4. $R^2$ is the coefficient of determination and $\beta$ is the gradient from the ordinary least squares regression of the x-axis variable onto SGW contribution to MLD.

[END OF MANUSCRIPT]